# Learning to screen Glaucoma like the ophthalmologists

Junde Wu[1], Huihui Fang[1], Fei Li[2], Huazhu Fu[3], Yanwu Xu[1]

Baidu, Healthcare Group[1]
State Key Laboratory of Ophthalmology, Zhongshan Ophthalmic Center, Sun Yat-sen University[2]
Institute of High Performance Computing (IHPC), Agency for Science, Technology and Research (A*STAR)[3]

-GAMMA Challenge is organized to encourage the AI models to screen the glaucoma from a combination of 2D fundus image and 3D optical coherence tomography volume, like the ophthalmologists.

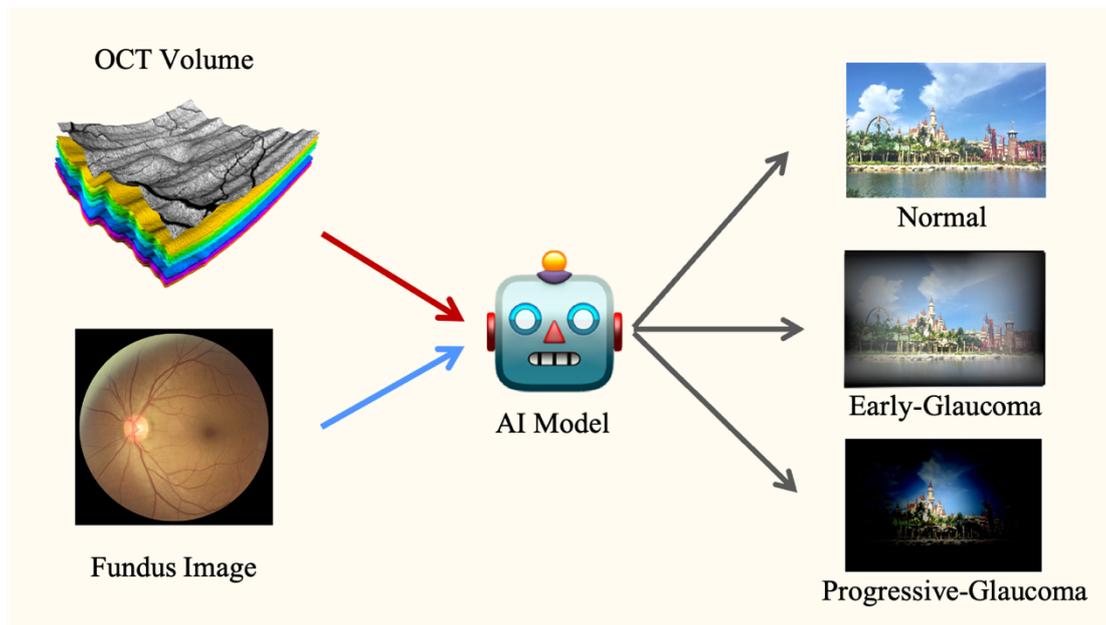

Figure 1. An illustration of the GAMMA Challenge. The goal of the challenge is to predict normal, early-glaucoma or progressive-glaucoma from fundus-OCT pairs using AI models.

Glaucoma is a chronic neuro-degenerative condition that is one of the world's leading causes of irreversible blindness[7], which is characterized by the damage of the optic nerve head (ONH) caused by a high intra-ocular pressure (IOP). Color fundus photography and Optical Coherence Tomography (OCT) are two most effective non-invasive tools for the glaucoma screening at present [8]. The two modalities of data are complementary as OCT volume contains cross-sectional images of the retinal tissues and fundus image is the 2D retinal tissues projection on the imaging plane. Both two modalities of data have their prominent biomarkers to indicate glaucoma, such as the vertical cup-to-disc ratio (vCDR) of fundus images and retinal nerve fiber layer (RNFL) thickness of OCT volume. Clinically, ophthalmologists often recommend to take both of the screenings for a more accurate and reliable diagnosis [9]. Recent report also shows nearly 46.3% glaucoma cases would be ignored if using only fundus images or OCT volume [1].

Recently, Artificial Intelligence (AI) models have shown to be efficient and effective in automated glaucoma screening. Computer-aided diagnosis models can provide an objective prediction to mitigate the human individual bias and to save ophthalmologist substantial time. However, automated diagnosis algorithms are commonly developed on only the fundus images. At present, there are still few models that make use of both 2D fundus image and 3D OCT volume to predict the glaucoma suspect like ophthalmologists. This is primarily due to two reasons: a) there has no sufficient paired multi-modality data to develop and evaluate the AI models. b) this multi-modal fusion task is technically challenging due to the large discrepancy between the two modalities.

In order to establish the AI model mimicking the way ophthalmologists detect the glaucoma, we set up the Glaucoma grAding from Multi-Modality imAges (GAMMA) Challenge to encourage the AI models to screen the glaucoma based on both fundus image and OCT volume. The primary task of the challenge is to grade glaucoma from the pairs consisting of 2D fundus image and 3D OCT scanning volume, as shown in Figure 1. We released a dataset of 300 pairs of fundus image and 3D OCT volume, including three sets each with 100 images for training, validation, and testing. During the preliminary stage of the competition, we released a training set for the participants. Teams are able to submit their methods and check their performance on the validation set. The preliminary stage lasted 30 days from 20 Mar 2021 to 01 October 2021. Each team is given five times to submit the predictions of model per day. A total of 70 teams submitted 1272 valid results to the platform during the preliminary, and 10 teams, which with top ranking were selected to the final stage. Finally, 10 teams are ranked based on their performance on the test set. During the final stage, teams are not allowed to modify their models anymore.

For each pair of data, the participants will predict normal, early-glaucoma or progressive-glaucoma. We use weighted Cohen's kappa as the evaluation metric. Cohen's Kappa coefficient is calculated based on the confusion matrix, with the value between -1 and 1. kappa = 0 denotes random guess. Higher Kappa denotes higher prediction performance.

The results of the challenge show well-designed AI models achieve much better performance on multi-modality data against single-modality one. The top three teams in the competition are SmartDSP, Voxelcloud and EyeStar, who achieve 0.855, 0.850 and 0.847 kappa respectively. For the comparison, the baseline model established on fundus images achieves 0.673 kappa, and which established on OCT volume achieves 0.575 kappa. A simple multi-modality fusion baseline model, which concatenates the feature of two modalities of data without bells and whistles achieves 0.702 kappa. We can see a simple fusion of 2D fundus and 3D OCT knowledge in the AI model gains about 10% improvement against the single-modality ones. Another 10% improvement can gain from the advanced design of the AI model. The sensitivities of top-3 teams (0.959, 0.918 and 0.959) are considerably higher than reported sensitivities of junior ophthalmologists (0.694 to 0.862) [1]. It indicates the models have the potential to be applied in the real clinical scenario for the glaucoma screening.

We find the recipes of winning teams were surprisingly consistent with the ophthalmologists' clinical experience. SmartDSP used a unique ensemble strategy to combine the predictions of multiple trained models. They picked three models with the best accuracy on normal, early, and progressive cases, respectively, and then ensembled the results by the priorities of early, progressive and normal. The uncertain cases in all three models will be classified as early-glaucoma by default. These strategies are often applied by clinical experts in their decision making. Clinically, ophthalmologists tend to assume the case as positive at first, and then look for sufficient evidence to rule out this suspicion. In addition, when multiple experts disagree, this case will be considered suspected early glaucoma for further screening. Such a clinical strategy helps to reduce the missed diagnosis. VoxelCloud and EyeStar abstracted the features from the source using unique models. VoxelCloud found 3D network[2] can better capture the correlation of multiple slices in 3D OCT volume. EyeStar found DENet[3] which focuses on the optic disc and cup region of the fundus images helps to improve the performance. The features selected in these strategies are also closely related to the clinical biomarkers. The RNFL thickness and vCDR calculated from OCT multi-slices and fundus optic disc region are discriminative parameters for the glaucoma diagnosis.

Despite the strong performance obtained by the winning teams, there is still room for improvement. We note that the multi-modality data fusion strategy in GAMMA are very straightforward. Eight of ten top teams simply concatenated the abstracted features of fundus images and OCT volume for the discrimination. The main reason is that many advanced multimodal fusion techniques recently proposed are not effective to this task. The fusion of fundus image and OCT volume is very different from the common multimodal fusion tasks. At present, multimodal fusion algorithms are often proposed for image-language pairs, which is inapplicable to our case. In medical image processing, multimodal fusion techniques are often proposed for the Computed Tomography (CT) and Magnetic Resonance Imaging (MRI) pairs. These methods often rely on the spatial correlation in CT-MRI pairs which does not exist in fundus-OCT pair [4,5,6], and thus to be invalid on this task. To our knowledge, few multimodal fusion techniques can be directly adopted on fundus & OCT fusion task. This indicates that more specific multimodal fusion algorithms are required in this field.

Inspired by the combination usage of fundus images and OCT volume in glaucoma screening clinically, GAMMA Challenge is organized to encourage the development the multi-modal fusion AI models for the automated glaucoma screening. Results show that, just like the ophthalmologists, well-designed AI models perform better when multi-modality data is provided. In the future work, we will explore the possibility of further combining IOP measurement data and visual field test data to create an automated glaucoma diagnosis AI model in full accordance with the clinical glaucoma diagnosis criteria.